\documentclass[pra,final,twocolumn,amsmath,eqsecnum,amssymb,superscriptaddress,showpacs]{revtex4-1}
\usepackage{graphicx}
\usepackage{color}

\usepackage{epsfig}
\usepackage{amsmath}
\usepackage{amssymb}
\usepackage{bm}

\begin{document}

\title{Charge transfer model
for the electronic structure of layered ruthenates}

\author{     Krzysztof Ro\'sciszewski}
\affiliation{Marian Smoluchowski Institute of Physics, Jagiellonian University,
             prof. S. \L{}ojasiewicza 11, PL-30348 Krak\'ow, Poland }

\author{     Andrzej M. Ole\'s  }
\affiliation{Marian Smoluchowski Institute of Physics, Jagiellonian University,
             prof. S. \L{}ojasiewicza 11, PL-30348 Krak\'ow, Poland }
\affiliation{Max-Planck-Institut f\"ur Festk\"orperforschung,
             Heisenbergstrasse 1, D-70569 Stuttgart, Germany }

\date{\today}

\begin{abstract}
Motivated by the earlier experimental results and \textit{ab initio}
studies on the electronic structure of layered ruthenates
(Sr$_2$RuO$_4$ and Ca$_2$RuO$_4$) we introduce and investigate the
multiband $d-p$ charge transfer model describing a single RuO$_4$
layer, similar to the charge transfer model for a single CuO$_2$ plane
including apical oxygen orbitals in high $T_c$ cuprates. The present
model takes into account nearest-neighbor anisotropic
ruthenium-oxygen $d-p$ and oxygen-oxygen $p-p$ hopping elements,
crystal-field splittings and spin-orbit coupling. The intraorbital
Coulomb repulsion and Hund's exchange are defined not only at ruthenium
but also at oxygen ions. Our results demonstrate that the RuO$_4$ layer
cannot be regarded to be a pure ruthenium $t_{2g}$ system. We examine a
different scenario in which ruthenium $e_g$ orbitals are partly
occupied and highlight the significance of oxygen orbitals. We point
out that the predictions of an idealized model based on ionic
configuration (with $n_0=4+4\times 6=28$ electrons per RuO$_4$ unit) do
not agree with the experimental facts for Sr$_2$RuO$_4$ which support
our finding that the electron number in the $d-p$ states is
significantly smaller. In fact, we find the electron occupation of $d$
and $p$ orbitals for a single RuO$_4$ unit $n=28-x$, being smaller by
at least 1--1.5 electrons from that in the ionic model and
corresponding to self-doping with $x\simeq 1.5$.
\end{abstract}


\maketitle

\section{Introduction}
\label{intro}

The description of the electronic states of transition metal oxides with
partly filled $d$ orbitals is not an easy task, and one usually looks
for some simplifications. Models for manganites are complex as they
involve partly filled $t_{2g}$ and $e_g$ orbitals \cite{Feh04}. Systems
with partly filled $t_{2g}$ orbitals and empty $e_g$ orbitals, such as
for titanates \cite{Pav05} or vanadates \cite{Sol06} or with completely
filled $t_{2g}$ and partly filled $e_g$ orbitals as in cuprates
\cite{Arr09} or nickelates \cite{Rei05} are much easier to investigate.
Such systems can be realized if the gap separating $t_{2g}$ from $e_g$
orbital states (induced by crystal-field effects) is sufficiently large.

The transition-metal oxides with $4d$ ions are even more challenging as
there electron correlations are somewhat weaker and simultaneously
spin-orbit coupling plays an important role. Therefore one has to treat
$4d$ electrons as both itinerant and strongly correlated, in the
vicinity of a metal-insulator transition. The compound which belongs to
this class and received a lot of attention is Sr$_2$RuO$_4$ as it became
a candidate for a $p$-wave superconductor \cite{Mac03}. Recent progress
in photoemission technique made it possible to investigate the many-body
effects both in bulk and surface bands of Sr$_2$RuO$_4$ \cite{Zab12}.
Spin-orbital entangled states have been seen recently in spin- and
angle-resolved photoemission spectroscopy \cite{Vee14}. Such states
arise in correlated transition metal oxides either on superexchange
bonds \cite{Ole12} or locally due to strong spin-orbit coupling at
transition-metal sites \cite{Jac09}. Indeed, these quantum effects play
an important role in Mott insulators with $4d$ ions \cite{Kha13} and in
these systems doped by $3d$ elements \cite{Hos12,Brz15}.

In Sr$_2$RuO$_4$ the experiments support the earlier implementation of
spin-orbit coupling within the local density approximation with Coulomb
interaction $U$ treated in LDA+$U$ approach which show
that both Coulomb $U$ and the spin-orbit coupling are necessary for a
correct description of the Fermi surface in Sr$_2$RuO$_4$ \cite{Hav08}.
The electronic structure of Sr$_2$RuO$_4$ was extensively studied in
the past \cite{Mac96} and it was established that the orbital
physics plays here an important role \cite{Fan05}. In Ca$_2$RuO$_4$
the bandwidth is smaller but the effects of spin-orbit coupling are
even more pronounced \cite{Liu11}.

Due to a rather large crystal-field splitting between $t_{2g}$ to $e_g$
levels $\sim 3.5$ eV in Ca$_2$RuO$_4$ and Sr$_2$RuO$_4$ \cite{Par01},
one might expect that these compounds are also purely $t_{2g}$ systems.
This picture was also supported by the earlier studies of electronic
structure in Ca$_{2-x}$Sr$_x$RuO$_4$ by photoemission \cite{Wan05}.
Orbital polarization is then helpful to understand the x-ray absorption
measurements and is of importance in describing the insulating state
of Ca$_{2-x}$Sr$_x$RuO$_4$ \cite{Miz04}. A~unique feature of
Ca$_{2-x}$Sr$_x$RuO$_4$ is that slight changes in lattice parameters
can induce drastic modifications of the character of their electronic
ground states --- Sr$_2$RuO$_4$ is metallic and superconducting at low
temperature \cite{Mac03}, whereas Ca$_2$RuO$_4$ is distorted and
undergoes a metal-insulator transition \cite{Nak97}. Interest in these
materials is also motivated by very unusual invar effect reported
recently \cite{Qi10} which suggests spin-orbital entanglement
\cite{Ole12} in the ground state.

Recently a simple tight-binding model was employed to investigate the
superconductivity in Sr$_2$RuO$_4$ \cite{Yan13}. The simplest $d-p$
model would just include three $t_{2g}$ orbitals per Ru and three $2p$
orbitals per O ion. However, there
are serious doubts about the validity of this simplified physical
picture. First, there are \textit{ab initio} cluster+embedding
computations by Kaplan and Soullard \cite{Kap07} who claim that the
$p$-orbital charges on oxygens (in Sr$_2$RuO$_4$) are not close to
formal 6.0 as they follow from the ionic model, but are instead close
to 5.0, and that in addition $e_g$ levels are partly occupied.
At the same time the transfer of
charge from any strontium ion (to ruthenium-oxide layer) is smaller
than a formal value of 2 electrons. Secondly, a similar  system, namely
CoO$_2$ layer was also believed to be pure $t_{2g}$ systems. However,
it was shown \cite{we} that $e_g$ orbitals (in a doped system) can
become very important, even in the absence of spin-orbit coupling.
We show below that the total electron density which follows
from the idealized ionic model with formal electronic charges does not
describe correctly the electronic structure of Sr$_2$RuO$_4$.
In contrast, the $d-p$ model with reduced electron density based on
\textit{ab initio} calculations \cite{Kap07} gives results
which agree with experiment.

To resolve the question about actual electron density within $e_g$
orbitals in Sr$_2$RuO$_4$ we constructed a multiband charge-transfer
model and performed unrestricted Hartree-Fock computations on a
finite RuO$_4$ cluster which contains $4\times 4$ Ru ions and
$4\times 4\times 4$ oxygen ions --- half of them located within the
same plane as Ru ions, while the second half belonging to the elongated
RuO$_6$ octahedra and surrounding the plane from above and below, being
in out-of-plane (apical) positions. We imposed cyclic boundary
conditions. The model involves (per a single RuO$_4$ unit) five $4d$
orbitals on Ru and $4\times 3$ oxygen $2p$ orbitals per unit cell
occupied by:
(i) $n_0=4+4\times 6$ electrons, according to the formal and
idealized ionic model;
(ii) the electron number lower by at least one electron, i.e.,
$n=3+4\times 6$ electrons or even smaller (we study below the case
of $n=2.5+4\times 6$ electrons), as found in Ref. \cite{Kap07}.

The paper is organized as follows. In Sec. \ref{sec:model} we introduce
the multiband model which includes all $4d$ states at ruthenium ions
and $2p$ states at oxygen ions. The parameters of the model are
specified in Sec. \ref{sec:para}. The Hartree-Fock approximation for the
Coulomb interactions is explained in Sec. \ref{sec:hf}, while in Sec.
\ref{sec:resu} we present
the results of numerical calculations and we introduce the concept of
self-doping with respect to the electron densities in the ionic model.
The paper is concluded with a short discussion and summary of the main
results in Sec. \ref{sec:summa}. In the Appendix we give the hopping
elements $d-p$ and $p-p$, respectively.

\section{Model Hamiltonian}
\label{sec:model}

In this section we introduce the $d-p$ charge-transfer Hamiltonian for
RuO$_4$ plane (such as realized in Sr$_2$RuO$_4$). It consists of
several parts,
\begin{equation}
{\cal H}= H_{\rm kin}+H_{\rm so}+ H_{\rm diag} +H_{\rm int}.
\label{model}
\end{equation}
The different terms in Eq. (\ref{model}) stand for the kinetic
energy ($H_{\rm kin}$), spin-orbit coupling ($H_{\rm so}$),
crystal-field splittings which are diagonal in the $\{4d,2p\}$ orbital
basis ($H_{\rm diag}$), and the intraatomic Coulomb interactions
($H_{\rm int}$) --- they all are explained below.

\subsection{Kinetic energy in hybridized $d-p$ bands}

The kinetic part of the Hamiltonian is:
\begin{equation}
H_{\rm kin}=   \sum_{ \{i, \mu; j, \nu\}, \sigma} \left(t_{ i, \mu; j,\nu}
 c^{\dagger}_{i,\mu,\sigma}  c_{j,\nu,\sigma}^{} + H.c.\right),
\end{equation}
where we employ a general notation, with $c_{j,\nu,\sigma}^{\dagger}$
standing for the creation of an electron at site $j$ in an orbital
$\nu$ with up and down spin, $\sigma=\uparrow,\downarrow$.
The model includes all $4d$ orbital states per Ru atom,
$\nu\in\{xy,yz,zx,x^2-y^2,3z^2-r^2\}$, and three $2p$ orbitals per
oxygen atom, $\nu\in\{p_x,p_y,p_z\}$. Alternatively, i.e., choosing a
more intuitive notation, we can write $d_{j,\nu,\sigma}^{\dagger}$ for
$d$ orbitals, while $p_{j,\nu,\sigma}^{\dagger}$ for $p$ orbitals.

The matrix $t_{i,\mu; j,\nu}$ is assumed to be non-zero only for
nearest neighbor ruthenium-oxygen $d-p$ pairs, and for nearest
neighbor oxygen-oxygen $p-p$ pairs. The next nearest hopping elements,
in particular direct ruthenium-ruthenium ones, and those between the
$p$ orbitals of neighboring apical oxygens are neglected.
The nonzero $t_{i,\mu; j,\nu}$ elements are listed in the Appendix.

\subsection{Spin-orbit coupling in layered ruthenates}

Formally, simplified spin-orbit part $H_{\rm so}$ of the Hamiltonian
Eq. (\ref{model}) has a similar mathematical structure to the kinetic
part $H_{\rm kin}$ \cite{Miz96a,Pol12,Mat13,Du13,Har13},
with $t^{so}_{\mu,\sigma;\nu,\sigma'}$ elements restricted to single
ruthenium sites,
\begin{eqnarray}
H_{\rm so}&=&  \sum_i  H_{\rm so}^{(i)}  \nonumber \\
&=&  \sum_i \left\{ \sum_{  \mu \neq \nu;\sigma, \sigma }
t^{so}_{\mu,\sigma;\nu,\sigma'}
 d^{\dagger}_{i, \mu, \sigma}  d_{i,\nu,\sigma'}^{} + \mathrm{H.c.}\right\},
\label{so-part}
\end{eqnarray}
where the summation runs only over ruthenium sites and where we
explicitly use $d_{i,\nu,\sigma'}^{\dagger}$ operators for the $4d$
orbitals at Ru sites. The derivation of spin-orbit coupling starts from
a single-site model. Using the $\{|i,\mu,\sigma\rangle\}$ basis one
evaluates the full matrix of scalar products,
$\langle i,\nu,\sigma'|\textbf{L}_i\cdot\textbf{S}_i|i,\mu,\sigma\rangle$,
of angular momentum $\textbf{L}_i$ with spin $\textbf{S}_i$ operator (at
site $i$). The individual single-site terms in $H_{so}^{(i)}$ are defined
by on-site hopping-like elements, $t_{{i\;\nu,\sigma';\mu,\sigma}}$
(between different spin and orbital states), and one arrives at the
matrix form in Eq. (\ref{so-part}).

As we use here the basis of real $4d$ orbitals (and not the spherical
harmonics) several spin-orbit elements turn out to be purely imaginary
(hence the hermitian Hamiltonian is not real but complex).
The elements of the matrix $t^{so}$ for a ruthenium site $i$ are the
following ones (for a similar result consult Fig. 6 in Ref. \cite{Pol12}),
\begin{widetext}
\begin{equation}
 H_{\rm so}^{(i)} =
 \frac{ \zeta}{2} \, \,  \left[
\begin{array}{rrrrrrrrrr}
 0  &   0  &   0   &  2i  &  0  &  0  &  1  & -i  &  0  &  0   \\
 0  &   0  &   i   &   0  &  0  & -1  &  0  &  0  &  i  &-\sqrt{3}i \\
 0  &  -i  &   0   &   0  &  0  &  i  &  0  &  0  & -1  & \sqrt{3}   \\
-2i &   0  &   0   &   0  &  0  &  0  &  i  &  1  &  0  &  0   \\
 0  &   0  &   0   &   0  &  0  &  0  &\sqrt{3}i&-\sqrt{3}&0&0 \\
 0  &  -1  & - i   &   0  &  0  &  0  &  0  &  0  & -2i &  0  \\
 1  &   0  &   0   &  -i  &-\sqrt{3}i&0 & 0 & -i  &  0  &  0  \\
 i  &   0  &   0   &   1  &-\sqrt{3} &0 &  i&  0  &  0  &  0  \\
 0  &  i   &  -1   &   0  &  0  & 2i  &  0  &  0  &  0  &  0  \\
 0  &\sqrt{3}i&\sqrt{3}&0 &  0  &  0  &  0  &  0  &  0  &  0  \\
\end{array} \right],
\end{equation}
\end{widetext}
where $\zeta$ is the spin-orbit coupling parameter and where the
columns and rows are labeled in the following order:
\begin{eqnarray}
& &(xy\uparrow),(yz\uparrow),(zx\uparrow),(x^2-y^2\!\uparrow),
(3z^2-r^2\!\uparrow), \nonumber \\
& &(xy\downarrow),(yz\downarrow),(zx\downarrow),(x^2-y^2\!\downarrow),
(3z^2-r^2\!\downarrow). \nonumber
\end{eqnarray}

Note that the consequence of finite spin-orbit coupling $\zeta$ is
nonconservation of the $z$th component of the total spin and
therefore the obtained ground state wave function is not a product
of two Slater determinants for $\uparrow$- and $\downarrow$-spin.
In some cases the spin-orbit coupling $H_{\rm so}$ can be treated as a
minor perturbation and can be neglected when the average value of local
angular momentum (at site $i$) is quenched to zero due to suitably
strong crystal-field effects and low enough local symmetry (reduced
by nearest neighbor atoms). Then, the spin-orbit Hamiltonian can
contribute to total energy only as second-order correction. Such a
reasoning however allows one to make only qualitative predictions.

An explicit treatment of spin-orbit coupling causes some difficulties.
It is likely that the true ground states are not homogeneous in space,
e.g. involving spin spirals or other micro-modulations, thus they may
be considered intractable within Hartree-Fock computations for $d-p$
clusters (they have too many order parameters to converge).
One can only hope that these micro-modulations are of secondary
importance. We assume this scenario and to have a tractable model we
decided to use a simplified approach. Namely, we break the symmetry
along natural quantization axis which is the $z$-th axis
(perpendicular to RuO$_4$ layer). We emphasize that the averages of
local spin components aligned parallel to the $(a,b)$ plane are assumed
to be zero. To present this assumption in a more transparent way
we can write down the formula for a local spin-flip,
$S_{i,\mu}^+ =  d_{i,\mu,\uparrow}^\dagger d_{i,\mu,\downarrow}^{}$,
and a similar one for $S_{i,\mu}^-$. Thus the requirements that
$\langle S_{i,\mu}^+\rangle=\langle S_{i,\mu}^-\rangle=0$
are equivalent to setting to zero the following order parameters with
opposite spins,
$\langle d^{\dagger}_{i,\mu,\sigma}d_{i,\mu,-\sigma}^{}\rangle=0$.

In Sec. \ref{sec:resu} we report a study of charge space-homogeneous
solutions. All occupation numbers, i.e., primary order parameters, are
assumed to be the same for equivalent ruthenium ions and similar for
oxygen ions. When studying the possibility of antiferromagnetism there
are two sublattices --- thus the number of order parameters doubles.
Looking for charge space-homogeneous ground states can be considered a
simplification but from another point of view it can be treated as a
consequence of strong long-range interionic electrostatic interactions.
These interactions are not explicitly included in the model
(\ref{model}) at present. We remark that long-range interionic
interactions cannot be easily incorporated into a typical $d-p$ model
but instead they can be accounted for by requiring that individual
ionic charges are space-homogeneous in accordance with crystal symmetry.

\subsection{Crystal-field splittings}

Let us now return to the $d-p$ Hamiltonian (\ref{model}). The next part
of the model $H_{\rm diag}$ is diagonal in the orbital basis and depends
only on electron number operators. It takes into account the effects of
crystal field and the difference of reference orbital energies (here we
employ the electron notation),
\begin{equation}
\Delta=\varepsilon_d-\varepsilon_p,
\label{Delta}
\end{equation}
between $d$ and $p$ orbitals, both for empty states without the Hartree
terms which follow from $H_{\rm int}$.
Below we fix the reference energy $\varepsilon_d$ for $d$ orbitals to
zero, hence we use only $\varepsilon_p$ as a parameter and write:
\begin{eqnarray}
H_{\rm diag}  &=&
\sum_{i,\mu=x,y,z;\sigma}
\varepsilon_p  p^\dagger_{i,\mu,\sigma}  p_{i,\mu,\sigma}^{} \nonumber \\
&+&\sum_{i,\mu=xy,yz,... ;\sigma}
f^{cr}_{\mu,\sigma} d^\dagger_{i,\mu,\sigma} d_{i,\mu,\sigma}^{}.
\end{eqnarray}
Here the first sum is restricted to oxygen sites, while the second one
runs over ruthenium sites.

What concerns the value of $\varepsilon_p$, it could be different for
in-plane and for out-of-plane (apical) oxygens. According to the
earlier studies \cite{Ogu95,Noc99} this difference can be as large as
1.5 eV. But such a large value follows from a simplified procedure of
fitting electronic bands to the LDA results. In the framework of the
present $d-p$ model one should expect much smaller difference, if any.
The zero difference was assumed in computations performed in Ref.
\cite{Sug13} and we also adopt this value. Here we remark that our
test Hartree-Fock computations performed using a big 1.5 eV splitting
give large differences in charge occupation between in-plane and
apical oxygens, in disagreement with the results of the population
analysis in Ref. \cite{Kap07}. This choice is indeed unrealistic as
in addition one obtains only ferromagnetic ground states, see below.

Let us return again to $H_{\rm diag}$. The vector containing the
elements of $\{f^{cr}_{\mu,\sigma}\}$ can be expressed as
\begin{equation}
f^{cr} = \frac{1}{3}\,D_1 \,\left[
\begin{array}{r} 2\\-1\\-1\\0\\0\\2\\-1\\-1\\0\\0 \end{array} \right]
+ D_2 \, \left[
\begin{array}{r} 0\\0\\0\\1\\0\\0\\0\\0\\1\\0 \end{array} \right]
+ D_3 \, \left[
\begin{array}{r} 0\\0\\0\\0\\1\\0\\0\\0\\0\\1 \end{array} \right].
\end{equation}
It includes the orbital splittings of $4d$ orbitals in the tetragonal
crystal field. The constant $D_1$ serves as a crude estimate of the
splitting between the orbital $xy$ and the orbital doublet $\{yz,zx\}$,
i.e., when taking into account only $H_{\rm diag}$ while neglecting
the remaining parts of the full Hamiltonian,. For $D_1<0$ (like in
Ca$_2$RuO$_4$) in the presence of a tetragonal distortion of RuO$_6$
octahedra the $xy$ orbital has a lower energy and is occupied (in the
atomic configuration) by two electrons, while $yz$ and $zx$ are
occupied by one electron each --- then the energy gain is $D_1$, i.e.,
such an occupation pattern is more stable.
For $D_1>0$ (like in Sr$_2$RuO$_4$)
the doublet $\{yz,zx\}$ has a lower energy and is more stable. The
parameters $D_2$ and $D_3$ follow from the estimates of the splitting
between $t_{2g}$ and $e_g$ orbitals. They do not influence the ground
state energy if only $t_{2g}$ orbitals are occupied, while there is a
punishment (by the value of $D_2$ or $D_3$) for each electron
occupying $x^2-y^2$ or $3z^2-r^2$ level respectively. These parameters
$\{D_2,D_3\}$ are much larger than $D_1$ in the case of Sr$_2$RuO$_4$
and the difference between them is rather small.

Jahn-Teller part can be neglected in the Hamiltonian (\ref{model}).
The exception is the elongation of bonds between ruthenium and apical
oxygens which could be considered as a frozen global (static)
$Q_3$ Jahn-Teller distortion \cite{Feh04} but it is much simpler to
include it by a proper renormalization of the crystal-field splittings.
Note that in the other compound Ca$_2$RuO$_4$ Jahn-Teller effects are
large (see for example Ref. \cite{Nak97}) and as a result the symmetry
of the corresponding RuO$_4$ plane is significantly lowered.

\subsection{Local Coulomb interactions}

The last part of the multiband $d-p$ Hamiltonian $H_{\rm int}$ stands
for strong on-site Coulomb interactions. For the $d$ orbitals at
ruthenium sites it includes Hubbard intraorbital repulsion $U_d$,
Hund's exchange $J^d_{\mu\nu}$ and pair hopping also given by
$J^d_{\mu\nu}$,
\begin{widetext}
\begin{eqnarray}
H^d_{\rm int}&=&
  U_d \sum_{i, \mu}  n_{i, \mu, \uparrow} n_{i, \mu, \downarrow}
+\frac{1}{2}\sum_{i,\mu\neq\nu}
\left(U_d-\frac{5}{2}J^d_{\mu\nu}\right)n_{i,\mu}n_{i,\nu}\nonumber  \\
&-& \sum_{i,\mu\neq\nu} J^d_{\mu\nu}\,\mathbf{S}_{i,\mu}\cdot\mathbf{S}_{i,\nu}
+  \sum_{i,\mu\neq\nu} J^d_{\mu\nu}\,
d^\dagger_{i,\mu, \uparrow} d^\dagger_{i,\mu, \downarrow}
d_{i,\nu, \downarrow}^{} d_{i,\nu, \uparrow}^{}.
\label{hubbard-intra}
\end{eqnarray}
Here $J^d_{\mu\nu}$ is the tensor of Hund's on-site interorbital
exchange elements for $d$ orbitals which can be expressed using Racah
parameters $B$ and $C$ \cite{Ole05,Gri71} (see also Table I given by
Horsch in Ref. \cite{Hor07}). Note that we sum twice over each pair
$\{\mu\nu\})$ of orbitals in Eq. (\ref{hubbard-intra}). Importance of
local Coulomb interactions has been recognized in several model
studies \cite{Mario,Beh12}. In particular, strong correlations which
originate from Hund's coupling have been studied \cite{Mra11} and
it was also suggested that this coupling supports the triplet
superconductivity \cite{Spa01}.

The anisotropy between different Hund's exchange elements
$\{J^d_{\mu\nu}\}$ may be neglected as long as one may limit oneself
to the orbitals of the same symmetry, i.e., either to $t_{2g}$ or to
$e_g$ orbitals \cite{Ole12}. For convenience, we rewrite equation
(\ref{hubbard-intra}) as follows
\begin{eqnarray}
H^d_{\rm int}&=&
  U_d \sum_{i, \mu}  n_{i,\mu, \uparrow} n_{i,\mu, \downarrow}
+\frac{1}{2}\sum_{i,\mu\neq\nu,\sigma} \left( U_d-3J^d_{\mu\nu}\right)
 n_{i,\mu,\sigma} n_{i,\nu,\sigma}
 + \frac{1}{2} \sum_{i,\mu\neq\nu,\sigma} \left(U_d-2J^d_{\mu\nu}\right)
 n_{i\mu, \sigma}n_{i\nu,-\sigma}   \nonumber\\
&-& \sum_{i,\mu\neq\nu} J^d_{\mu\nu}\,
d^\dagger_{i,\mu \uparrow} d_{i\mu, \downarrow}^{}
d^\dagger_{i,\nu \downarrow} d_{i\nu, \uparrow}^{}
+  \sum_{i,\mu\neq\nu} J^d_{\mu\nu}\,
d^\dagger_{i,\mu, \uparrow} d^\dagger_{i, \mu, \downarrow}
d_{i, \nu, \downarrow}^{}d_{i, \nu, \uparrow}^{}\,.
\label{hubbard2-intra}
\end{eqnarray}
Similarly, local Coulomb interactions at oxygen sites
(for $2p$ orbitals) are given by
\begin{eqnarray}
H^p_{\rm int}&=&
  U_p \sum_{i, \mu}  n_{i, \mu, \uparrow} n_{i, \mu, \downarrow}
+\frac{1}{2}\sum_{i,\mu\neq\nu,\sigma} \left(U_p-3J^p_{\rm H}\right)
 n_{i, \mu,\sigma} n_{i, \nu,\sigma}
 + \frac{1}{2} \sum_{i,\mu\neq\nu,\sigma} \left(U_p-2J^p_{\rm H}\right)
 n_{i, \mu, \sigma}n_{i, \nu,-\sigma}  \nonumber\\
&-&  \sum_{i,\mu\neq\nu} J^p_{\rm H}\,
p^\dagger_{i, \mu, \uparrow} p_{i, \mu, \downarrow}^{}
p^\dagger_{i, \nu, \downarrow} p_{i, \nu, \uparrow}^{}
+\sum_{i,\mu\neq\nu} J^p_{\rm H}\,
p^\dagger_{i, \mu, \uparrow}p^\dagger_{i, \mu, \downarrow}
p_{i, \nu, \downarrow}^{}p_{i, \nu, \uparrow}^{}\,,
\label{hubbard3-intra}
\end{eqnarray}
\end{widetext}
where all off-diagonal elements $J^p_{\mu\nu}$ are equal as they connect
the orbitals of the same symmetry, i.e.,
$J^p_{\mu\nu}\equiv J^p_{\rm H}$. Up to now this latter part
($H^p_{\rm int}$) was neglected in the majority of studies, i.e., they
assume $U_p=J^p_{\rm H}=0$. As a compensation some effective and
appropriate modification of the charge-transfer energy has to be
performed. Indeed, it has been shown for La$_{2-x}$Sr$_x$CuO$_4$
cuprates employing constrained local density approach that the
estimation of $U_p$ is very sensitive to the assumed value of $\Delta$
\cite{Hyb89} --- the value of $\Delta=2.0$ eV yields $U_p\sim 8$ eV,
while $\Delta=4.0$ eV yields $U_p\sim 4$ eV. One notes that there is a
roughly linear dependence of $U_p$ on $\Delta$ to reproduce a constant
value of the charge-transfer gap in Hartree-Fock \cite{Ole91}; at the
same time the estimated value of $U_d$ remains more or less constant.

Unfortunately, not much is known about the real value of $\Delta$
(\ref{Delta}) for different compounds. Estimates based on LDA results
for Sr$_2$RuO$_4$ gave the value $\Delta\sim 1.5$ eV \cite{Noc99}. This
value is much lower than $\Delta\sim 8$ eV used in Ref. \cite{Sug13}
(where also $U_p =0$ was assumed). All these estimations come from
different fitting procedures employing numerous simplifying assumptions.
One has to realize that $\Delta$ (\ref{Delta}) is not an
\textit{ab initio} like value in the framework od $d-p$ model but
rather some effective value which better should be treated as a free
parameter (the $d-p$ model itself is an effective model, and definitely
not an \textit{ab initio} model). Thus in the following we will vary
the value of $\Delta$ from 0 to 6.0 eV. We do not consider negative
values of $\Delta$ as their consequence are greatly overcharged
$d$-shells. We also remark that it has been realized nowadays that
effective models with only $d$ orbitals are insufficient and many
papers treat electronic oxygen degrees of freedom explicitly. For
cuprates this subject has a long history and realistic $d-p$ models
were studied for CuO$_3$ chains \cite{Ole91,Bie13} and for CuO$_2$
planes \cite{Arr09,Eme87,Jef92}. The oxygen orbitals are also of
importance for ruthenates such as in Sr$_2$RuO$_4$ \cite{Yos09} and
for other correlated oxides \cite{Bal94,Kor96,Vau12}.

\section{Setting the Hamiltonian parameters}
\label{sec:para}

\subsection{Previous studies of the charge-transfer model}

The effective $d-p$ model requires a choice of a number of explicitly
included parameters. In the extreme case when only $4d$ orbitals are
used in a tight-binding (semiempirical) model \cite{Yan13} the
parameters are very different from the cases where Coulomb interactions
are treated in the Hartree-Fock approximation. Here we adopt in-plane
hopping elements $(pd\sigma)$, $(pd\pi)$, $(pp\sigma)$ and $(pp\pi)$
used in \cite{Sug13}: $-3.4$, 1.53, 0.6, $-0.15$ (all in eV). The
out-of-plane hoppings (involving the apical oxygens) were scaled using
the formulae from the book by Harrison \cite{Har05}: $-2.6$, 1.167,
0.559, $-0.140$ (all in eV). A similar value of $(pd\pi)=1.5$ eV for
the in-plane hopping was reported earlier by Oguchi \cite{Ogu95} who
applied the tight-binding formulae to fit the LDA electronic
structure. Different estimations for $(pd\pi)$ are quite close to one
another; they read as follows:
(i) 1.0 eV  in Ref. \cite{Mis00};
(ii) $\sim$ 1 eV as used by the group of Fujimori's \cite{Har13};
(iii) 0.85 eV in Ref. \cite{Noc99}.
Let us note that for cuprates frequently accepted values for
$(pd\pi)$ are 0.75 eV in Ref. \cite{Hyb89} and 0.9 eV in Ref.
\cite{Arr09}.

The choice of the Coulomb elements is rather difficult due to their
considerable screening in the oxides which is however less efficient in
$4d$ systems \cite{Vau12}. For the intraorbital Coulomb repulsion $U_d$
at ruthenium sites the value of $\sim 3$ eV is most frequently
used \cite{Sug13,Pch07,Miz01,Har13}. We fix here $U_d=3.1$ eV
following Ref. \cite{Pch07}. We remark that we cannot follow popular
estimations made by Okamoto and Millis \cite{Oka04} and by Liebsch
\cite{Lie03} as they apply to effective models featuring only Ru
sites renormalized by the hybridization with oxygen orbitals.

Hund's exchange elements are less screened than intraorbital Coulomb
elements and close to their atomic values (see for example Ref.
\cite{marel88}). For Hund's exchange $J_{\rm H}^d$ between two $t_{2g}$
electrons various estimates range from 0.5 eV up to 0.8 eV:
(i)   0.5 eV in Refs. \cite{Sug13,Miz01};
(ii)  0.6-0.8 eV in Ref. \cite{Kho03};
(iii) 0.7 eV in Ref. \cite{Pch07};
(iv)  0.8 eV in Ref. \cite{Par01}.
We will use $J_{\rm H}^d=0.7$ eV. Moreover, for the sake of fixing
precisely Hund's coupling tensor elements $J^d_{\mu,\nu}$ we use Table I
from Ref. \cite{Hor07} and in addition we use an empiric formula
$C\simeq 4B$ for Racah parameters. With this \textit{Ansatz} for a pure
$t_{2g}$ system $J^d_{\rm H}=3B+C\approx 7B$ and $B=0.1$ eV \cite{Ole12}.
This determines the $J^d_{\mu,\nu}$ elements when $e_g$ levels are not
empty (using again the entries from Table I in Ref. \cite{Hor07}).

For the intraorbital Coulomb repulsion at oxygen sites $U_p$
(in ruthenates) again not much is known and it was neglected in several
studies. In Ref. \cite{Kho03} this element is estimated to be $U_p=4-6$
eV. In cuprates the available data are more abundant: $U_p$ is 4.5 eV
in Ref. \cite{Arr09}, $\sim$ 4 eV in Refs. \cite{Hyb89,Gra92}; while
several possibilities were also given (all in the range 3-8 eV) with
6 eV indicated by some experimental data \cite{Esk91}. We use below
$U_p=4.4$ eV. For Hund's coupling at oxygen ions the values
$J_{\rm H}^p=0.6-0.8$ eV were suggested \cite{Kho03}, while Grant and
McMahan computations in cuprates yield $J_{\rm H}^p=0.8$ eV \cite{Gra92}.
Following these estimations, we use below $J_{\rm H}^p=0.8$ eV. Note
that the corresponding tensor $J^p_{\mu,\nu}=J^p_{\rm H}$ has also the
same entries for all off-diagonal elements.

The spin-orbit coupling $\zeta$ on Ru sites is usually assumed to be in
range from 0.10 to 0.17 eV \cite{Vee14,Sug13,Har13,Mat13,Dai08}. Here
we follow the most recent experiments which suggest that this coupling
is in the middle of this range and take the value $\zeta=0.13$ eV
\cite{Vee14}.

To complete the set of the Hamiltonian parameters we have to provide
estimates for the crystal-field splittings. The splitting between
$xy$ and $\{yz,zx\}$ orbital levels is estimated as:
(i)  $\sim 1$ eV in Ref. \cite{Noc99};
(ii) $\sim 0.3$ eV in Refs. \cite{Mis00,Har13};
(iii) 0.1 eV in Ref. \cite{Pch07}.
We choose the value 0.1 eV, i.e., we trust the reliable expertise
presented in Ref. \cite{Pch07}.
The splitting between $t_{2g}$ and $e_g$ orbital levels is 3 eV
according to \cite{Par01} (in Ca$_2$RuO$_4$) and up to 3.5 eV
\cite{Har13}, while the splitting of 0.8 eV between $e_g$ orbital
levels was assumed \cite{Har13}. We accept these values in the
parameter set employed in the present calculations (see Table I).

\begin{table}[t!]
\caption{Parameters of the Hamiltonian (\ref{model}) (all in eV) used for
Hartree-Fock calculations. For the hopping integrals we adopt the values
from \cite{Sug13}. Below we present only representative in-plane Slater
integrals $(pd\pi)$ and $(pp\pi)$. Out-of-plane integrals are obtained
by applying Harrison scaling. Note that during computations we are
setting the value of $\varepsilon_d$ to be zero as the reference energy.
The value of $\Delta$ ($-\varepsilon_p$) is fixed to be the same for
in-plane and for apical oxygens and is studied in the range
$\Delta\in [0.0,6.0]$ eV.}
\begin{ruledtabular}
\begin{tabular}{ccccccccccc}
  $U_d$ & $J_{\rm H}^d $ & $U_p$ & $J_{\rm H}^p$ & $\zeta$ & $D_1$
        & $D_2$    & $D_3$ & $(pd\pi)$ &  $(pp\pi)$ \\ \hline
   3.1  &    0.7   &  4.4  &   0.8     &    0.13
&  0.10 &    4.3   &  3.5  &   1.53    &  $-0.15$   \\
\end{tabular}
\end{ruledtabular}
\label{tab:para}
\end{table}

Looking at the parameters, one important remark is proper. Namely the
three different values:
(i) the absolute value of parameter $(pd\sigma)$ which is involved in
hopping processes from $t_{2g}$ to $e_g$ orbitals (see Table V in the
Appendix);
(ii) the value of $U_d$; and also
(iii) the splitting between $t_{2g}$ and $e_g$ are all close to 3 eV.
In other words --- the splitting between $t_{2g}$ and $e_g$ does not
seem to be large enough to justify the expectation that $e_g$ levels
are almost empty.

\subsection{Motivation by earlier ab initio results}

Let us repeat that in the majority of the papers it is being assumed
that $e_g$ levels are entirely empty. However, in the present paper we
make an attempt to determine the electron densities in $e_g$ orbitals
and to provide a realistic estimate of the charge on oxygens. We remind
the reader that we are motivated by the \textit{ab initio} computations
performed on a small cluster+embedding by Kaplan and Soullard
\cite{Kap07}. We take the liberty to repeat, once more, these results
as they are really important for a proper understanding of the
electronic structure of Sr$_2$RuO$_4$:
(i) the $p$ orbital charge on oxygens (in Sr$_2$RuO$_4$) is not
formal 6.0 but is closer to 5.0 (oxygen $s$ orbitals are also not
fully occupied);
(ii) the occupation on $d$ orbitals is close to 6 but $e_g$ levels
are partly occupied;
(iii)~charges on strontium ions are not formal 2$^+$ but rather
$\sim 1.6^+$.

Direct mapping of the \textit{ab initio} results to the $d-p$ model is
not possible (as the $d-p$ model neglects the valence $s$ orbitals).
However, it seems clear that the formal (idealized) ionic model with
6 electrons occupying $p$ levels of each oxygen and 4 electrons
occupying $d$ levels of each Ru ion does not apply to the realistic
Sr$_2$RuO$_4$. For the sake of convenience let as take a convention
and introduce the \textit{self-doping} $x$ for a single RuO$_4$ unit
with respect to the formal idealized model ($x=0$) while for the real
substance we shall consider finite self-doping values such as $1.0$,
$1.25$ and $1.5$ ($n=28-x$, here we use such simple numbers so as
the self-doping translates into integer electron number for the entire
cluster).

\section{The unrestricted Hartree-Fock approximation}
\label{sec:hf}

\subsection{The self-consistent Hartree-Fock problem}

We use the unrestricted Hartree-Fock approximation to investigate the
model (\ref{model}) for Sr$_2$RuO$_4$. The technical implementation is
the same as described in Refs. \cite{Miz96a,Sug13,Miz01}. Namely,
the local Coulomb interaction Hamiltonian $H_{intra}$ is replaced by
Hartree-Fock mean-field terms. To give an example, the term according
to a common interpretation of Wick, Bloch, and de Dominicis theorem
the term $U_d \sum_{i, \mu}  n_{i,\mu, \uparrow} n_{i,\mu, \downarrow}$
can be replaced (for Hartree-Fock computations) with one-electron
operators and double counting correction terms,
\begin{widetext}
\begin{eqnarray}
\sum_{i, \mu}  n_{i,\mu, \uparrow} n_{i,\mu, \downarrow}
&\simeq&  \sum_{i,\mu} \left(
\langle d^\dagger_{i,\mu,\uparrow}d_{i,\mu,\uparrow}^{}\rangle
        d^\dagger_{i,\mu,\downarrow}d_{i,\mu,\downarrow}^{}  +
 d^\dagger_{i,\mu,\uparrow}d_{i,\mu,\uparrow}^{}
\langle d^\dagger_{i,\mu,\downarrow}d_{i,\mu,\downarrow}^{}\rangle
\right) \nonumber \\
&-&\!\sum_{i,\mu}  \left(
\langle d^\dagger_{i,\mu,\uparrow}  d_{i,\mu,\downarrow}^{}\rangle
        d^\dagger_{i,\mu,\downarrow}d_{i,\mu,\uparrow}^{}
+ d^\dagger_{i,\mu,\uparrow}  d_{i,\mu,\downarrow}^{}
\langle d^\dagger_{i,\mu,\downarrow}d_{i,\mu,\uparrow}^{}\rangle
\right) \nonumber \\
&-&\!\sum_{i,\mu}   \left(
\langle d^\dagger_{i,\mu,\uparrow}  d_{i,\mu,\uparrow}^{}\rangle
\langle d^\dagger_{i,\mu,\downarrow}d_{i,\mu,\downarrow}^{}\rangle
- \langle d^\dagger_{i,\mu,\uparrow}d_{i,\mu,\downarrow}^{}\rangle
\langle d^\dagger_{i,\mu,\downarrow} d_{i,\mu,\uparrow}^{}\rangle\right).
\end{eqnarray}
\end{widetext}
Note that the terms with superconducting correlations are ignored in
the above. Note also that standard mean-field decoupling usually
ignores spin-flip terms (second line) \cite{Miz96a,Sug13}.
We remind that spin-flip terms do appear in spin-orbit part of the
Hamiltonian (\ref{model}), see Eq. (\ref{so-part}), and have to
be included here on equal footing as the mean-field terms.

The averages
$\langle d^\dagger_{i,\mu,\uparrow}d_{i,\mu,\uparrow}^{}\rangle$
and other similar ones can be treated as order parameters. At the
beginning some initial values (a guess) have to be assigned to them.
During Hartree-Fock iterations the order parameters are recalculated
self-consistently until convergence. When decoupling all the terms in
$H_{\rm int}$ (\ref{hubbard-intra}) one finds that the complete set of
order parameters is as follows:
\begin{eqnarray*}
\langle d^\dagger_{i,\mu,\uparrow}  d_{i,\mu,\uparrow}^{}\rangle, \;\;
\langle d^\dagger_{i,\mu,\downarrow}d_{i,\mu,\downarrow}^{}\rangle, \;\;
\langle d^\dagger_{i,\mu,\uparrow}  d_{i,\mu,\downarrow}^{}\rangle,    \;\;
\langle d^\dagger_{i,\mu,\downarrow}d_{i,\mu,\uparrow}^{}\rangle, \\
\langle d^\dagger_{i,\mu,\uparrow}  d_{i,\nu,\uparrow}^{}\rangle,    \;\;
\langle d^\dagger_{i,\mu,\downarrow}d_{i,\nu,\downarrow}^{}\rangle,  \;\;
\langle d^\dagger_{i,\mu,\uparrow}  d_{i,\nu,\downarrow}^{}\rangle, \;\;
\langle d^\dagger_{i,\mu,\downarrow}d_{i,\nu,\uparrow}^{}\rangle,  \\
\end{eqnarray*}
where $\mu\neq\nu$. The off-diagonal elements ($\mu\neq\nu$) are of
crucial importance, particularly in the present case when finite
spin-orbit coupling induces their finite values. A similar set of order
parameters has to be considered for the $p$ oxygen orbitals.

\subsection{Hartree-Fock calculations for Sr$_2$RuO$_4$}

As we mentioned earlier, we are interested only in charge-homogeneous
solutions, i.e., homogeneity concerns only primary order parameters
($\mu = \nu$, i.e., charge occupations) but not off-diagonal
($\mu\neq\nu$) order parameters. Thus primary order parameters are
translationally invariant according to assumed lattice symmetry. In
addition, the symmetry is broken along the $z$-th axis and the order
parameters
$\{\langle d^\dagger_{i,\mu,\uparrow}d_{i,\mu,\downarrow}^{}\rangle\}$
are fixed to be zero. Finally, the four-fold symmetry is imposed so
that the occupation of $p_x$ and $p_y$ oxygen orbitals is the same,
and also of $yz$ and $zx$ orbitals for Ru ions. Altogether we have got
7 independent primary order parameters (per RuO$_4$ unit) when looking
for paramagnetic ground state and 15 for ferromagnetic
(or antiferromagnetic) ground states. These numbers are large enough to
expect troubles with the Hartree-Fock convergence and indeed this is
the case. The regular convergence is found for a very limited set of
the Hamiltonian parameters (for example for situations when the oxygen
occupations are very close to 6 as was the case in Ref. \cite{Sug13}).
A typical situation for our computations is that Hartree-Fock
iterations do not converge but oscillate (in a two-cycle) instead.

The standard remedy for poor convergence is the so-called dumping
technique, but here it failed, unfortunately. We had to resort to
quantum chemistry technique called level shifting \cite{Sou73}. It is
based on replacing the Hartree-Fock Hamiltonian by a different
Hamiltonian --- the one with the identical eigenvectors (one particle
eigenfunctions) and with identical occupied eigenenergies. The original
eigenenergies of virtual states are all uniformly shifted upwards by a
fixed constant value. Thus if we apply the shift say by 1 eV, then the
HOMO-LUMO gap (the gap between highest occupied and lowest unoccupied
eigenstate) we obtain will be artificially enlarged exactly by 1 eV.

\begin{figure}[t!]
\vskip 1.1cm
\begin{center}
\includegraphics[width=8cm]{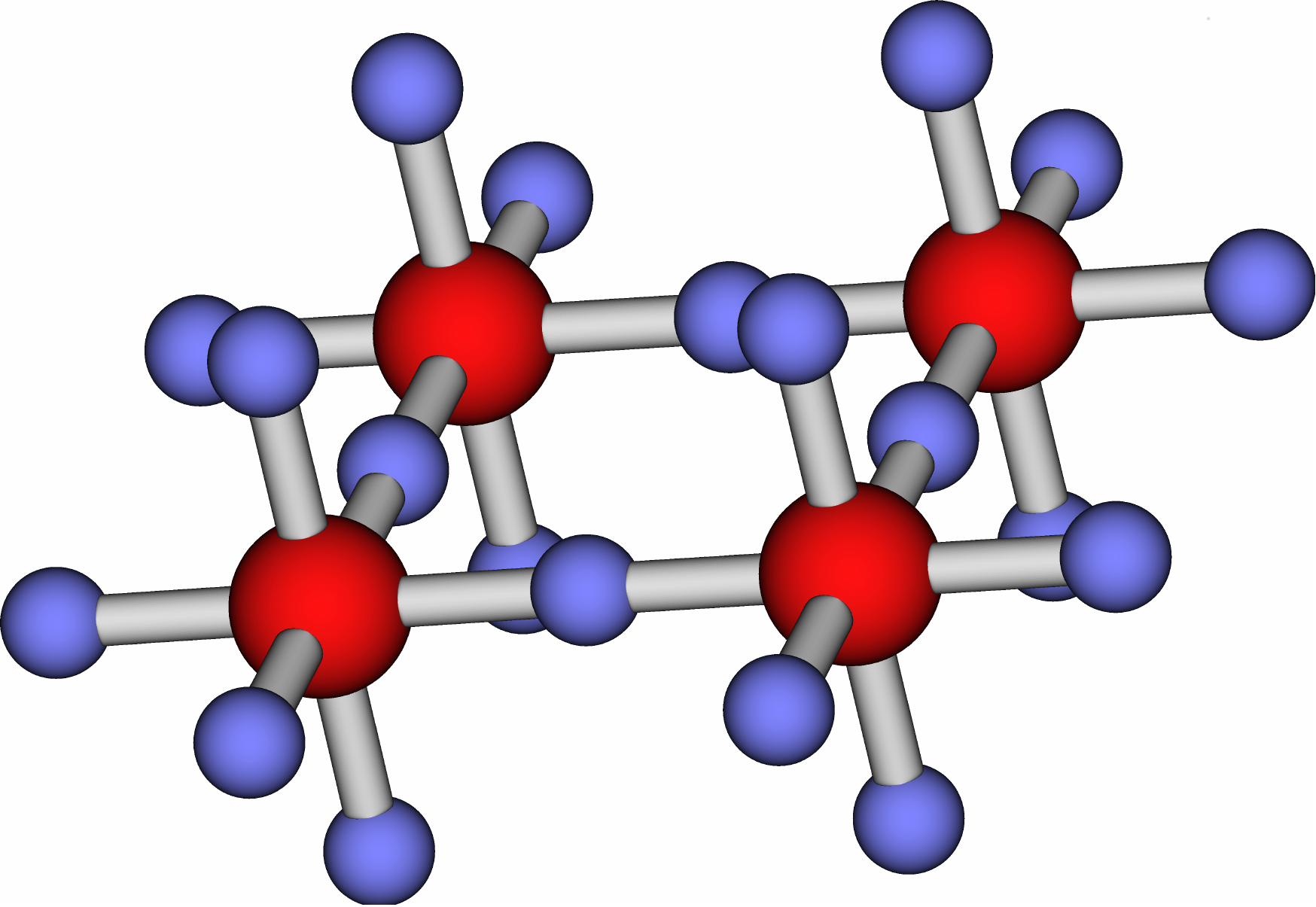}
\end{center}
\caption{A fragment of the studied CuO$_4$ cluster with four Ru ions
shown by dark (red) big spheres and surrounding them O ions shown by
smaller gray (dark blue) spheres. The calculations in Sec. V were
performed for a $4\times 4$ RuO$_4$ periodic supercell.
}
\label{clu}
\end{figure}

When applying virtual level shifting technique we can obtain
valuable information. First case is when the HOMO-LUMO splitting
(after correcting for the shift) is negative
(for a few different shifts and a few different starting conditions).
Then the single-determinant Hartree-Fock ground state we obtain is
probably not correct and single-determinant description of the ground
state is not possible at all. Multi-configuration Hartree-Fock is
required in such a situation instead (and let us remind that the
multi-configuration Hartree-Fock still did not mature enough to be a
standard working tool in solid state physics). Second case is when
HOMO-LUMO gap we obtain is zero (or very close to zero). Then the ground
state identification is questionable. However, the probability that such
an identification is correct and that the ground state is conducting
can be substantial. The probability of correct identification can be
further enhanced when performing numerous extra computations: if we
obtain the same identification for different shifts and different
starting Hartree-Fock conditions, the result is accepted. Finally for
the cases with positive HOMO-LUMO gap we usually had no problems.

The computations were performed for a periodic $4\times 4$ RuO$_4$
supercell with its fragment shown in Fig. \ref{clu} (for each
particular set of Hamiltonian parameters). They were repeated many
times for different starting conditions (different starting charge
occupations) and shifts. The state with the lowest Hartree-Fock energy
was then identified as the ground state. Numerous runs were necessary
as the Hartree-Fock convergence provides many different metastable
states (metastable, i.e., only local but not a global minima of the
energy). Typical number of runs should be
large enough --- in some situations more than a hundred. For this very
reason the detailed investigation of the phase diagram is too expensive
(even for such a small cluster size as we use). Still we performed as
many computations for as many different sets of Hamiltonian parameters
so as to be sure about the general trends occurring on the phase diagram
which are presented in Sec. V.

\section{Numerical results}
\label{sec:resu}

Numerical studies of the multiband $d-p$ model (\ref{model}) require
not only the parameters which were fixed in Sec. \ref{sec:para}, but
also an assumption concerning the total electron number per unit cell.
We consider below two different scenarios:
(i) the formal ionic model with $n_0=4+4\times 6=28$ electrons per
RuO$_4$ unit, and
(ii) the model with a smaller total number of $n=28-x$ electrons,
where we investigated a few representative values of
\textit{self-doping} $x=1.0,1.25,1.5$.
Thereby we concentrate on the most important results obtained for a
realistic value of Coulomb interaction within oxygen $2p$ orbitals,
$U_p=4.4$ eV, within the framework of these two different scenarios.
As we shall show below, these two situations require quite different
values of Coulomb parameter $U_p$ at oxygen $2p$ orbitals.

\begin{figure}[b!]
\begin{center}
\includegraphics[width=8cm]{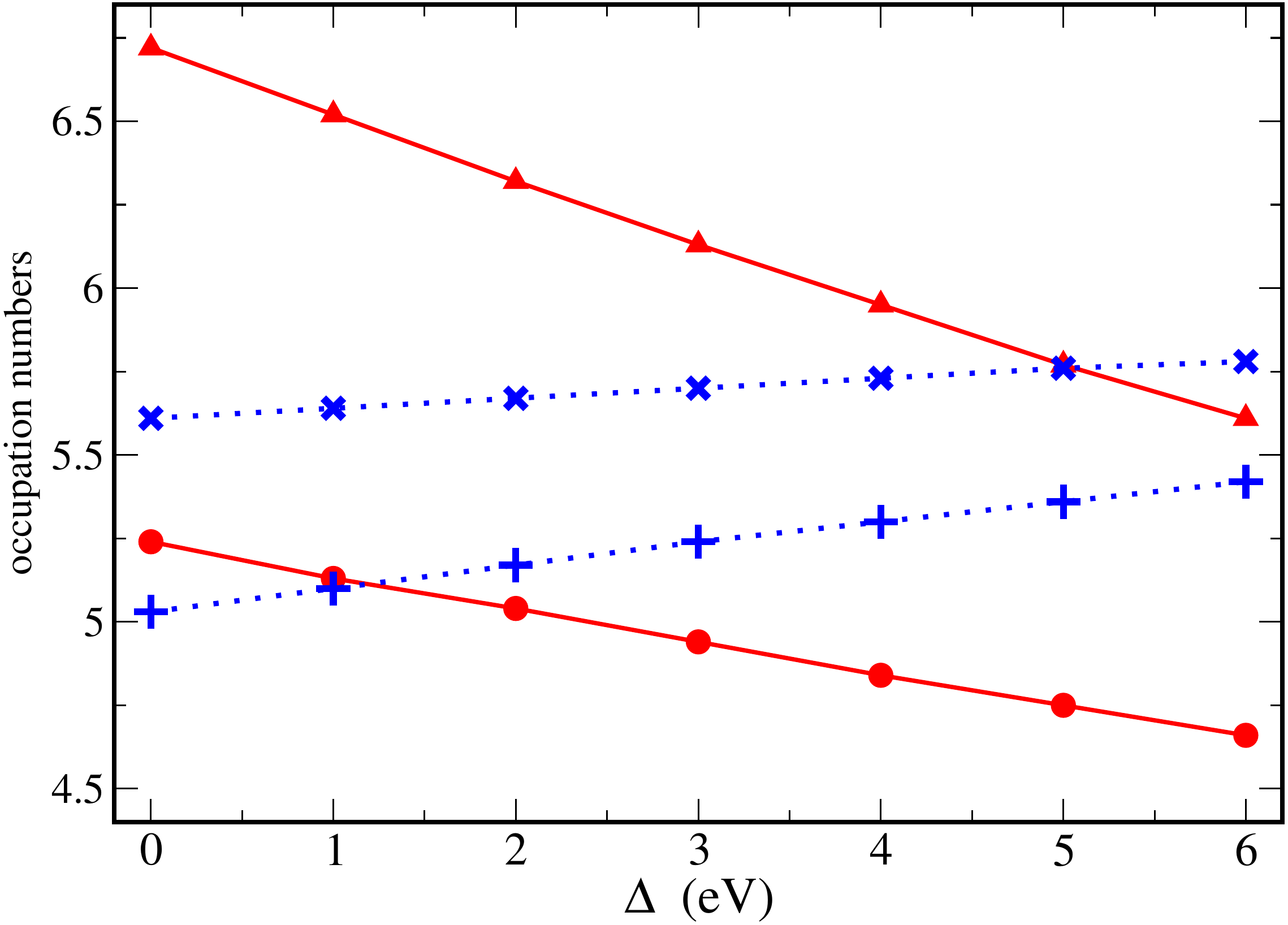}
\end{center}
\caption{Electron densities in the multiband model (\ref{model}) for
increasing $\Delta$ ($\Delta\equiv-\varepsilon_p$) obtained
in the formal ionic model (with $n_0=28$ electrons per RuO$_4$ unit):
at Ru ions (solid lines) and at O ions (dashed lines).
Data points show electron densities within
$t_{2g}$ orbitals ($n_{t2g}$, circles),
all $4d$ orbitals ($n_{4d}$, triangles),
$2p$ orbitals at in-plane oxygens ($n_{p\parallel}$, $+$), and
$2p$ orbitals at apical oxygens ($n_{p\perp}$, $\times$).
The obtained ground state is ferromagnetic.}
\label{fig2}
\end{figure}

\subsection{First scenario: Formal ionic model}

Taking the ionic model as a starting configuration for the Hartree-Fock
iterations, we assume that each oxygen O$^{2-}$ ion has 6 electrons
within $2p$ orbitals and each Ru$^{4+}$ ion has 4 electrons within
$4d$ orbitals. The RuO$_4$ unit has a negative charge $Q=4e$ which is
compensated by two Sr$^{2+}$ ions considered only as electron donors
to the plane of RuO$_4$ units. This charge distribution (assumed on
start of Hartree-Fock iterations) is however unstable and the electrons
quickly redistribute along the iteration process due to $d-p$
hybridization. The final charge distribution is shown in Fig. \ref{fig2}
and in the upper part of Table II. As expected, the total
($n_{4d}$) and partial ($n_{t2g}$) electron densities at Ru ions
increase with increasing value of $\varepsilon_p$
(i.e., decrease with increasing value of $\Delta$)
which follows from electron transfer from O to Ru ions.

\begin{table}[t!]
\caption{The ground state obtained in the formal ionic model with
$n_0=28$ ($x=0$) electrons and in the realistic model with self-doping
($x>0$): HOMO-LUMO gaps $G$ (eV), the total magnetic moment
$m_{\rm tot}$ per RuO$_4$ unit, and the magnetic moment at Ru ions
$m_{\rm Ru}$ obtained for several values of $\Delta$ (eV).
Note that when the total magnetization $m_{\rm tot}$ is large, it is
mainly due to the magnetization at oxygen ions.}
\begin{ruledtabular}
\begin{tabular}{ccccc}
      & \multicolumn{2}{c}{energies (eV)} & \multicolumn{2}{c}{magnetizations} \\
 $x$  & $\Delta$ & $G$ & $m_{\rm tot}$ & $m_{\rm Ru}$ \\ \hline
 0.00 & 0.0 & 0.10 & 1.00 & 0.34 \\
      & 1.0 & 0.29 & 0.75 & 0.32 \\
      & 2.0 & 0.39 & 0.75 & 0.36 \\
      & 3.0 & 0.43 & 0.75 & 0.40 \\
      & 4.0 & 0.46 & 0.75 & 0.45 \\ \hline
 1.00 & 0.0 & 0.27 & 1.25 & 0.30 \\
      & 1.0 & 0.10 & 0.24 & 0.09 \\
      & 2.0 & 0.30 & 0.24 & 0.11 \\
      & 3.0 & 0.33 & 0.24 & 0.12 \\
      & 4.0 & 0.35 & 0.24 & 0.14 \\ \hline
 1.25 & 0.0 &$\sim 0$&1.12& 0.26 \\
      & 1.0 &$\sim 0$&0.12& 0.11 \\
      & 2.0 & 0.04 & 0.11 & 0.05 \\
      & 3.0 & 0.04 & 0.11 & 0.06 \\
      & 4.0 & 0.04 & 0.11 & 0.06 \\   \hline
 1.50 & 0.0 & 0.22 & 0.20 & 0.18 \\
      & 1.0 & 0.21 & 0.00 & 0.00 \\
      & 2.0 & 0.26 & 0.00 & 0.00 \\
      & 3.0 & 0.38 & 0.00 & 0.00 \\
      & 4.0 & 0.51 & 0.00 & 0.00 \\
\end{tabular}
\end{ruledtabular}
\label{tab2}
\end{table}

The ground state (in the entire range of the investigated values of
$\varepsilon_p$) is ferromagnetic and insulating, which is surprising
and does not agree with \textit{ab initio} calculations
\cite{Hav08}. It could be argued that Hund's exchange is strong
enough to polarize the $4d$ electrons if $t_{2g}$ orbitals are well
away from half filling ($n_{t2g}>4.7$ for the considered range of
$\varepsilon_p$), in spite of large $d-p$ hybridization. The
instability towards ferromagnetism competes here with $d-p$
hybridization and therefore the magnetic moment per RuO$_4$ unit is
small. The nonmagnetic ground state is metastable, and has a higher
energy by about 0.8 eV (per RuO$_4$ unit). Also, the occupation
patterns for the obtained ground states do not agree with the
\textit{ab initio} data \cite{Kap07}. It is well known that
Sr$_2$RuO$_4$ has a paramagnetic and metallic ground state. Thus,
the model (\ref{model}) for the adopted parameter values and within
the first scenario is clearly not realistic enough for Sr$_2$RuO$_4$.

\subsection{Second scenario: Realistic self-doping model}

\begin{figure}[b!]
\begin{center}
\includegraphics[width=8cm]{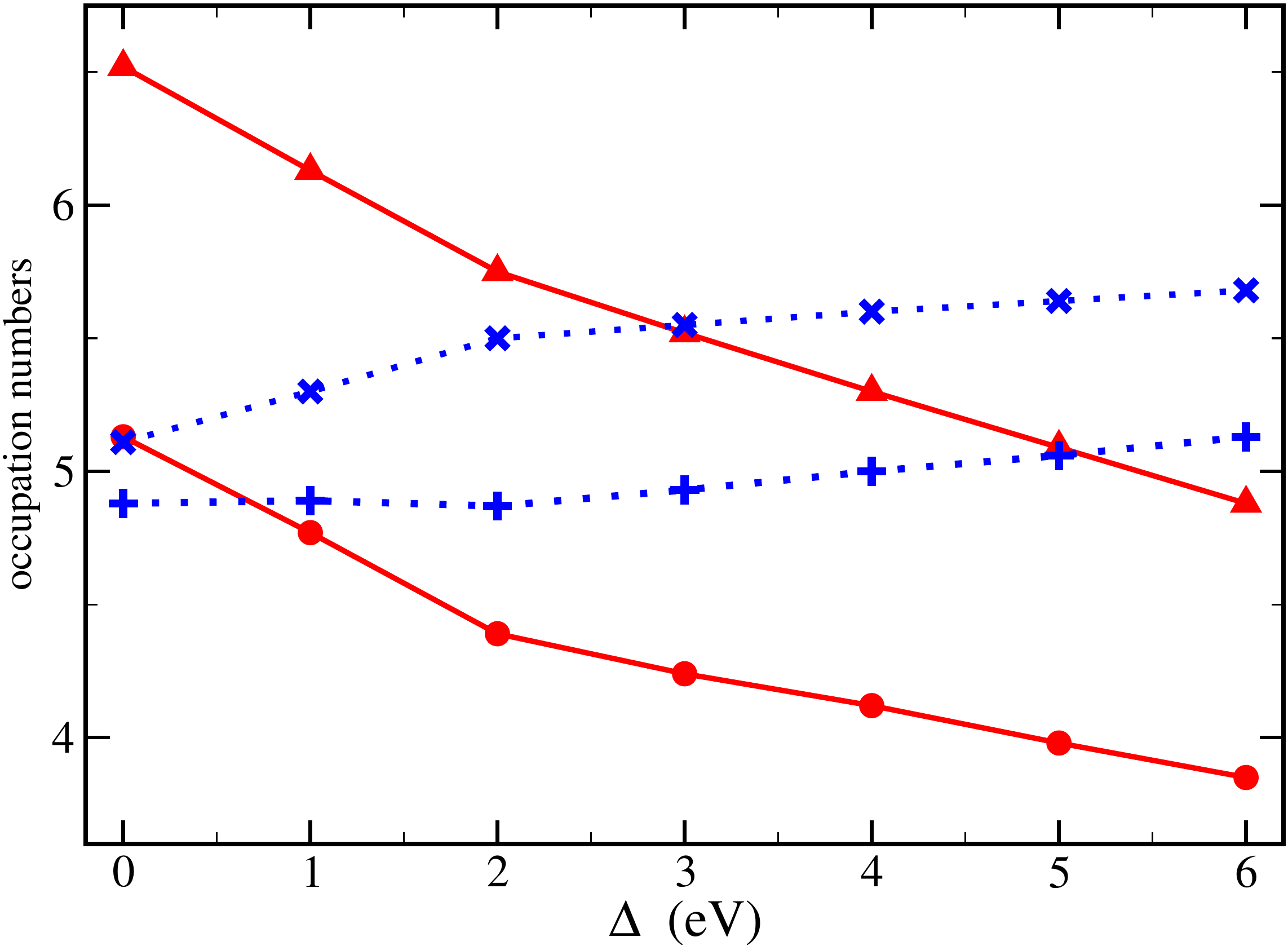}
\end{center}
\caption{Total electron densities in the realistic $d-p$ multiband
model based on \textit{ab initio} calculations for increasing
$\Delta$ ($\Delta=-\varepsilon_p$) for Ru($t_{2g}$) orbitals
(solid line, circles), Ru($4d$) orbitals (solid line, triangles)
and O($2p$) orbitals (dashed lines). There are $(4\times 6+2.5)$
electrons per a single RuO$_4$ unit, corresponding to self-doping by
$x=1.5$ electrons. The ground state is nonmagnetic for
$\Delta>1.0$ eV, and weakly ferromagnetic for $\Delta\leq 1.0$ eV.
The change of magnetic order is
responsible for the change of the slope of the lines.
}
\label{fig3}
\end{figure}

In the second scenario we follow the results presented in
\cite{Kap07} and we assume a reduced total number of electrons per
RuO$_4$ unit. Taking the total electron number $n=28-x$ with $x>0$
this corresponds to finite hole \textit{self-doping}, and we study
here $x=1.00,1.25,1.50$. For the missing electrons the smaller
(than in the ideal-ionic-model) transfer of valence $4s$ electrons
from Sr sites is mainly responsible ($5s$ valence electrons on Ru,
neglected in the $d-p$ model, have also some
minor influence). The corresponding electron densities obtained for
$x=1.50$ are shown in Fig. \ref{fig3} and in Table II.

The difference between the two density distributions shown in Figs. 2
and 3 is mainly visible in the electron densities at Ru ions. At finite
self-doping of $x=1.5$ the total electron density within $4d$ orbitals
is close to $n_{4d}=5$ for a value $\Delta \simeq 5.0$ eV, while it is
close to $n_{4d}=5.5$ for the same value of $\Delta$
in the ionic model. The electron density at in-plane oxygens is also
somewhat reduced in the former case. These orbitals are influenced
stronger by the self-doping as they are hybridized with the $t_{2g}$
orbitals at the central Ru ion in each RuO$_4$ unit, and provide also
bonding between these units.

Our calculations demonstrate that the metallic nonmagnetic state
observed in Sr$_2$RuO$_4$ is realized when the density of electrons
within Ru $t_{2g}$ orbitals is not close to half filling, i.e.,
$n_{t2g}<4.5$. Such a nomagnetic state is found in the present
realistic model with self-doping when of $2p$ energy $\Delta>1.0$ eV.
On the contrary, when $x\in[1.25,1.50]$ and $\Delta\in[0.0,1.0]$ eV,
the ground state is ferromagnetic. Metallic ferromagnetism is here
possible due to large values of the Stoner parameter
$I\simeq U_d+2J_H^d$ for partly filled $t_{2g}$ orbitals \cite{Sto90}
which is farther enhanced by partly occupied $e_g$ orbitals.
Effectively $I$ is enhanced here when the density of $4d$ electrons is
increased and $e_g$ orbitals are also partly filled as we have seen in
the ionic model.

\begin{table}[t!]
\caption{Hartree-Fock energy $E_{\rm HF}$, the HOMO-LUMO gap $G$,
electron densities within $t_{2g}$ orbitals ($n_{t2g}$), for all
$4d$ orbitals ($n_{4d}$), oxygen $2p$ orbitals at in-plane oxygens
($n_{p\parallel}$), and oxygen $2p$ orbitals at oxygens in apical
positions ($n_{p\perp}$), and total magnetic moments $m_{\rm tot}$
(per RuO$_4$ unit), for a few selected ground states with the total
electron number $n=28-x$ per RuO$_4$ unit.
Parameter: $\Delta=1.0$ eV.
}
\begin{ruledtabular}
\begin{tabular}{cccccccc}
      & \multicolumn{2}{c}{energies} (eV) & \multicolumn{5}{c}{electron densities} \\
  $x$ & $E_{\rm HF}$ & $G$ & $n_{t2g}$ & $n_{4d}$
                           & $n_{p\parallel}$ & $n_{p\perp}$ & $m_{\rm tot}$ \\ \hline
 1.00 & 111.049     & 0.24 & 4.78 & 6.19 & 4.95 & 5.45 & 0.10 \\
 1.25 & 107.838     & 0.12 & 4.85 & 6.21 & 4.95 & 5.32 & 0.00 \\
 1.50 & 104.663     & 0.00 & 4.77 & 6.13 & 4.89 & 5.30 & 0.21 \\
\end{tabular}
\end{ruledtabular}
\label{tab3}
\end{table}

The most interesting data obtained in our Hartree-Fock calculations for
the realistic model are presented in Table III. Here we identify the
region of phase diagram where both the HOMO-LUMO gap $G$ is small or
vanishes and the ground state is close to a transition from a
nonmagnetic to ferromagnetic one. In this regime one finds large
electron density within $e_g$ orbitals accompanied by rather strong
reduction of electron density at the oxygen ions in RuO$_2$ planes.
The density at these ions $n_{p\parallel}$ varies from $4.95$ to $4.89$
when $x\in[1.0,1.5]$, i.e., each oxygen ion contains one hole and
is rather close to the O$^{1-}$ ionic state. These results of our
computations agree well enough with the results of Ref. \cite{Kap07}.
Indeed, the charge on oxygen ions, in particular the ones lying within
RuO$_2$ planes, is close to 5 and not to formal 6 electrons per ion.
This demonstrates
the metallic character of the electronic structure in Sr$_2$RuO$_4$.
We emphasize that the occupations which follow from the present model
are close to those reported in the \textit{ab initio} investigation
\cite{Kap07} --- some representative examples of the occupations
obtained in the Hartree-Fock calculations are shown in Table III
(see also Fig. 3).

\subsection{Importance of Coulomb interactions at oxygen ions}

We have verified that actual electron densities within $2p$ orbitals
are rather sensitive to the used Coulomb interaction parameters at
oxygen ions. To obtain unphysical density of $n_{p\parallel}\simeq 6$
at in-plane oxygen ions (as in the formal ionic model) one must require
that $U_p=0$ and/or $\varepsilon_p$ must be very large negative, i.e.,
the charge-transfer gap $\Delta$ has to be very large.
This is confirmed by all test computations performed varying the
values of $U_p$ and $\varepsilon_p$.

\begin{table}[t!]
\caption{Hartree-Fock results for ferromagnetic ground states obtained
using three different miltiband models
in absence of electron interactions at oxygen ions ($U_p=J^p_{\rm H}=0$):
total electron densities within $t_{2g}$ orbitals ($n_{t2g}$), for all
$4d$ orbitals ($n_{4d}$), oxygen $2p$ orbitals at in-plane oxygens
($n_{p\parallel}$), and oxygen $2p$ orbitals at oxygens in apical
positions ($n_{p\perp}$), total magnetization $m_{\rm tot}$ per RuO$_4$
unit, HOMO-LUMO gap $G$ (eV), and spin-orbit contribution per single
Ru $\langle H_{so}^{(i)}\rangle$ (eV). In the third segment all
HOMO-LUMO gaps $G$ are zero (within 1 meV accuracy), thus the reliable
identification of the ferromagnetic ground state is not possible
(but probable).}
\begin{ruledtabular}
\begin{tabular}{cccccccccc}
        &   & \multicolumn{3}{c}{energies (eV)} & \multicolumn{5}{c}{electron densities} \\
     model  & $x$ & $\Delta$ & $G$ & $\langle H_{so}^{(i)}\rangle$
& $n_{t2g}$ & $n_{eg}$ & $n_{p\parallel}$ & $n_{p\perp}$ & $m_{\rm tot}$ \\ \hline
     I, all $4d$ & 0.0 & $ 0.0$ & 0.39 & $-0.076$ & 4.19 & 0.51 & 5.75 & 5.91 &  1.0  \\
                 &     & $ 3.0$ & 0.34 & $-0.083$ & 4.12 & 0.40 & 5.81 & 5.93 &  1.0  \\
                 &     & $ 6.0$ & 0.16 & $-0.088$ & 4.08 & 0.31 & 5.85 & 5.95 &  1.0  \\ \hline
 I, only $t_{2g}$& 0.0 &   0.0  & 0.37 & $-0.043$ & 4.24 &  --  & 5.91 & 5.98 &  1.0  \\
                 &     & $ 3.0$ & 0.35 & $-0.049$ & 4.14 &  --  & 5.94 & 5.99 &  1.0  \\
                 &     & $ 6.0$ & 0.20 & $-0.052$ & 4.09 &  --  & 5.96 & 5.99 &  1.0  \\ \hline
 realistic       & 1.5 &   0.0  &$\sim 0$&$-0.026$& 2.90 & 0.62 & 5.63 & 5.86 & 1.25  \\
                 &     & $ 3.0$ &$\sim 0$&$-0.025$& 2.76 & 0.50 & 5.73 & 5.91 & 1.25  \\
                 &     & $ 6.0$ &$\sim 0$&$-0.025$& 2.67 & 0.37 & 5.80 & 5.93 & 1.25
\label{tab4}
\end{tabular}
\end{ruledtabular}
\end{table}

Using the set of Hamiltonian parameters from Table 1 but setting
$U_p=J^p_{\rm H}=0$ (while keeping other parameters unchanged) we
performed additional Hartree-Fock computations to investigate
importance of Coulomb repulsion at oxygen ions. These calculations
gave very different electron density distributions than those obtained
before for the same values of $\Delta$ and total electron density $n$,
but with finite $U_p$ and $J^p_{\rm H}$. As shown in Table IV, one
finds large electron densities at in-plane oxygens,
$n_{p\parallel}>5.75$, and only ferromagnetic ground states in the
entire range of $\varepsilon_p\in[-6,0]$. It may be considered quite
unexpected that the $2p$ oxygen orbitals are almost completely filled
then even at $\varepsilon_p=0$. The almost ionic state O$^{2-}$ is here
a consequence of Coulomb repulsion at Ru ions which blocks electron
redistribution due to hybridization. It is also surprising that the
same Hartree-Fock energy is obtained for ferromagnetic and for
antiferromagnetic ground states.
Note that for the pure $t_{2g}$ model (when setting $D_2\gg 1$ and
$D_3\gg 1$) and for large $\Delta=6.0$ eV (when $t_{2g}$
occupation number is equal to formal $n_{t2g}=4$) the nonmagnetic
ground state has a higher energy by $\sim 0.8$ eV than the
ferromagnetic one.

\section{Discussion and summary}
\label{sec:summa}

Altogether the results of the presented Hartree-Fock computations are
too complex to be fully conclusive, but nevertheless this study uncovers
several important facts concerning the modeling of ruthenium oxides by
the multiband charge-transfer model. First of all, the $d-p$ model with
a minimal basis set consisting of $\{4d,2p\}$ orbitals is a useful tool
for investigating the electronic structure when the effective electron
density within the considered basis set is established in agreement
with the experimentally observed ground states. We have found that a
significant electron charge is transferred \textit{beyond} the $d-p$
orbitals and thus the effective electron density within the RuO$_4$
units has to be reduced to $n=28-x$, with $x\in[1.0,1.5]$. This effect
is similar to the reduction of electron density in $d-p$ orbitals in
cuprates, where $4s$ orbitals at Cu ions are also partly occupied
\cite{Pav01}. Another reason responsible for this appreciably reduced
electron density could be a partial charge transfer from oxygen orbital
to the charged Sr ions in Sr$_2$RuO$_4$, suggesting that the ionic
picture with Sr$^{2+}$ ions transferring 2 electrons to the RuO$_4$
subsystem is oversimplified.

We have shown that Coulomb interaction effects at oxygen ions are very
important and have to be included in a realistic description of these
materials. Only then the hybridization effects are strong enough and
are able to overcome Hund's exchange at ruthenium ions which otherwise
induces metallic ferromagnetic state, contrary to the experimental
observations. We note however that ferromagnetic instability was
observed in Ca$_{2-x}$Sr$_x$RuO$_4$ systems where antiferromagnetic
interactions are also possible \cite{Nak03}. This is reminiscent of the
situation encountered in the Ca$_{1-x}$Sr$_x$RuO$_3$ perovskites, with
CaRuO$_3$ found to be on the verge of a ferromagnetic instability
\cite{Maz97}. Therefore, we suggest that further research in a model
including lattice distortions is required to establish the range of
stability of ferromagnetism in the layered Ca$_{2-x}$Sr$_x$RuO$_4$
systems, being to some extent also expected from the present results.

An extension of this model could be used for a similar modeling of the
electronic structure of Ca$_2$RuO$_4$, but this would also require
Jahn-Teller coupling to the lattice to describe correctly the lattice
distortions which accompany the insulating state. It remains a challenge
for the theory to establish whether the electron density within the
multiband $d-p$ model would not be increased by such an insulating state
and we suggest that the self-doping effect described here would be
concentration dependent in the Ca$_{2-x}$Sr$_x$RuO$_4$ compounds.

In summary, the most important consequence of both $d-p$ hybridization
and spin-orbit coupling is the
electron transfer from $t_{2g}$ to $e_g$ orbitals as our calculations
demonstrate that $e_g$ orbitals at Ru ions are partly occupied in the
realistic parameter regime. Thus, $e_g$ orbitals have to be included
in any realistic model for ruthenium oxides. This invalidates the
paradigm that ruthenium oxides are pure $t_{2g}$ systems. A second very
important effect is a significant reduction of the electron density
within oxygen orbitals from the values obtained in the ionic model,
which effectively corresponds to one hole per oxygen ion within RuO$_2$
planes. Finally, only when the above self-doping effect is included,
the nonmagnetic metallic state of Sr$_2$RuO$_4$ may be correctly described.

\acknowledgments

It is our pleasure to thank Atsushi Fujimori for insightful discussions.
We kindly acknowledge financial support by Narodowe Centrum Nauki (NCN, 
Polish National Science Center) under Project No. 2012/04/A/ST3/00331.

\begin{table}[t!]
\caption{The non-zero ruthenium-oxygen hopping elements in RuO$_4$
plane as obtained using Slater-Koster rules \cite{Har05,Sla54}.
$(pd\pi)$ and $(pd\sigma)$ are the appropriate Slater-Koster
interatomic integrals \cite{Sla54,Har05}. We assume that oxygen ions
belonging to the $i^{th}$ RuO$_6$ octahedron are reached by in-plane
vectors $\pm{\bf a}_1=\pm a(1,0,0)$, $\pm{\bf a}_2=\pm a(0,1,0)$, and
the apical oxygen positions by out-of-plane vectors,
$\pm{\bf a}_3=\pm b(0,0,1)$.
We use standard notation, with ${\bf R}={\bf R}_i-{\bf R }_j$ being the
vector for a nearest neighbor bond between ruthenium at site $i$ and
oxygen at site $j$, while the coordinates $(l,m,n)$ of ${\bf R}/R$
which stand for the direction cosines of the hopping.
}
\begin{ruledtabular}
\begin{tabular}{ccccc}
    $\bf{R}$      &  $(l,m,n)$  & $\nu$ &   $\mu$    &   $t_{j,\nu;i,\mu}$   \\ \hline
$\pm${\bf{a}}$_1$ & $(\pm 1,0,0)$ & $x$ & $x^2-y^2 $ & $l(\frac{\sqrt{3}}{2})(pd\sigma)$ \\
                  & $(\pm 1,0,0)$ & $x$ & $3z^2-r^2$ & $-l(\frac{1}{2})(pd\sigma)$   \\
                  & $(\pm 1,0,0)$ & $y$ & $xy$       & $l(pd\pi)$    \\
                  & $(\pm 1,0,0)$ & $z$ & $zx$       & $l(pd\pi)$    \\ \hline
$\pm${\bf{a}}$_2$ & $(0,\pm 1,0)$ & $y$ & $x^2-y^2$  & $-m(\frac{\sqrt{3}}{2})(pd\sigma)$ \\
                  & $(0,\pm 1,0)$ & $y$ & $3z^2-r^2$ & $-m(\frac{1}{2})(pd\sigma)$ \\
                  & $(0,\pm 1,0)$ & $x$ & $xy$       & $m(pd\pi)$    \\
                  & $(0,\pm 1,0)$ & $z$ & $yz$       & $m(pd\pi)$    \\ \hline
$\pm${\bf{a}}$_3$ & $(0,0,\pm 1)$ & $z$ & $3z^2-r^2$ & $n(pd\sigma)$ \\
                  & $(0,0,\pm 1)$ & $x$ & $xz$       & $n(pd\pi)$    \\
                  & $(0,0,\pm 1)$ & $y$ & $yz$       & $n(pd\pi)$    \\
\end{tabular}
\end{ruledtabular}
\end{table}

\appendix*

\section*{Appendix}

The non-zero ruthenium-oxygen $d-p$ and oxygen-oxygen $p-p$ hopping
elements obtained by using Slater-Koster rules \cite{Sla54} for the
lattice constant $a=1$ are presented in Tables V and VI, respectively.

\begin{table}[t!]
\caption{The non-zero oxygen-oxygen hopping elements in RuO$_4$ plane
as obtained using Slater-Koster rules \cite{Har05,Sla54}.
$(pp\pi)$ and $(pp\sigma)$ are the appropriate Slater-Koster
interatomic integrals \cite{Har05,Sla54}.
We use a standard notation as explained in Table V, with
symbols $(l,m,n)$ standing for the direction cosines of the hopping.
}
\begin{ruledtabular}
\begin{tabular}{ccccc}
      $p-p$ hopping             & $(l,m,n)$ & $\nu$ & $\mu$ & $t_{j,\nu;i,\mu}$   \\ \hline
$2p_{\parallel}-2p_{\parallel}$ & $(\pm\frac{\sqrt{2}}{2},\pm\frac{\sqrt{2}}{2},0)$
                          & $x$ & $x$& $(\frac{1}{2})\left[(pp\sigma)+(pp\pi)\right]$ \\
                                & $(\pm\frac{\sqrt{2}}{2},\pm\frac{\sqrt{2}}{2},0)$
                          & $y$ & $y$& $(\frac{1}{2})\left[(pp\sigma)+(pp\pi)\right]$ \\
                                & $(\pm\frac{\sqrt{2}}{2},\pm\frac{\sqrt{2}}{2},0)$
                          & $z$ & $z$& $(pp\pi)$            \\
                                & $(\pm\frac{\sqrt{2}}{2},\pm\frac{\sqrt{2}}{2},0)$
                          & $x$ & $y$ & $lm\left[(pp\sigma)-(pp\pi)\right]$ \\ \hline
$2p_{\parallel}-2p_{\perp}$ & $(\pm\frac{\sqrt{2}}{2},0,\pm\frac{\sqrt{2}}{2})$
                      & $x$ & $x$ & $(\frac{1}{2})\left[(pp\sigma)+(pp\pi)\right]$ \\
                            & $(\pm \frac{\sqrt{2}}{2},0,\pm\frac{\sqrt{2}}{2})$
                      & $y$ & $y$ & $(pp\pi)$ \\
                            & $(\pm\frac{\sqrt{2}}{2},0,\pm \frac{\sqrt{2}}{2})$
                      & $z$ & $z$& $(\frac{1}{2})\left[(pp\sigma)+(pp\pi)\right]$  \\
                            & $(\pm\frac{\sqrt{2}}{2},0,\pm\frac{\sqrt{2}}{2})$
                      & $x$ & $z$ & $ln\left[(pp\sigma)-(pp\pi)\right]$     \\ \hline
$2p_{\parallel}-2p_{\perp}$ & $(0,\pm\frac{\sqrt{2}}{2},\pm\frac{\sqrt{2}}{2})$
                      & $x$ & $x$ & $(pp\pi)$   \\
                            & $(0,\pm\frac{\sqrt{2}}{2},\pm\frac{\sqrt{2}}{2})$
                      & $y$ & $y$ & $(\frac{1}{2})\left[(pp\sigma)+(pp\pi)\right]$ \\
                            & $(0,\pm\frac{\sqrt{2}}{2},\pm\frac{\sqrt{2}}{2})$
                      & $y$ & $z$ & $mn\left[(pp\sigma)-(pp\pi)\right]$     \\
                            & $(0,\pm\frac{\sqrt{2}}{2},\pm\frac{\sqrt{2}}{2})$
                      & $z$ & $z$ & $(\frac{1}{2})\left[(pp\sigma)+(pp\pi)\right]$ \\
\end{tabular}
\end{ruledtabular}
\end{table}

\end{document}